\documentclass[a4paper,10pt]{article}

\usepackage[dvips]{graphicx}
\usepackage{epsfig}
\usepackage{graphicx}
\usepackage{subcaption}
\usepackage{amssymb}
\usepackage{amsmath}
\usepackage{textcomp}
\usepackage{cite}
\usepackage{subcaption}
\def\be{\begin{equation}}

\def\ee{\end{equation}}
\begin{document}
\title{Cosmic scenario of early Universe in context of Euclidean wormhole in dilatonic Einstein Gauss Bonnet gravity}
\author{ Gargi Biswas$^{}$\footnote{E-mail: biswasgb@gmail.com} and B Modak$^{}$\footnote{E-mail: bijanmodak@yahoo.co.in}~}

\maketitle
\noindent
\begin{center}
\noindent
Department of Physics, University of Kalyani, Kalyani, India-741235\\

\end{center}

\begin{abstract}
We present some wormhole configurations using analytic and numerical solutions of the Euclidean field equations considering Einstein Gauss Bonnet dilaton coupled interaction in the Robertson Walker metric.
Analytic solutions invoking slow-roll type approximation for some potentials yield identical wormholes in the early era. Eventually, in a later era, the universe enters to an expanding era much faster than usual inflation accompanied by oscillations with analytic continuation $\tau=it$. The numerical solutions of wormholes in terms of the scale factor $a(\tau)$ are almost identical for some standard potentials of the dilaton field, except for the evolution of potentials. The Hubble parameter and deceleration parameter obtained by curve fit of $a(\tau)$ of the numerical solution of wormhole yield an inflationary era away from the throat of the wormhole using $\tau=it$. A sharp decay of the potential is also observed.
 \end{abstract}
\section{Introduction}
A brief period of early inflation in GUT era invoking a large value of vacuum energy or cosmological constant \cite{gu:prd,li:pl,stei:prl,gu:prl,a:plb} is necessary to resolve some of the problems of standard cosmology. However, it fails to address the problems of initial cosmic singularity and vanishing of cosmological constant required for cosmic evolution \cite{pd:book}. Einstein's gravity is not fully attuned to resolve these issues appearing in a strong curvature regime in the early universe, rather we need quantization of gravity and matter fields. In this endeavour, higher-order curvature terms appear naturally from the fundamental of Physics as re-normalization terms in quantum theory in curved spacetime \cite{stell:prd,qft:bd}. At this juncture, modified theories \cite{capp:prd,harko:prd,lobo:fR} of gravity appear as an alternative theory of gravitation, which can accommodate not only inflation \cite{a:plb} but also late-time acceleration \cite{capgara:cqg,capfran:grg,bm:aspace} of the universe. Kaluza-Klein gravity \cite{sahdev:plb,wsson:AJ,Dzhu:prd}, low energy Heterotic string theory \cite{met:string,sloan:string} etc. built on the basis of using extra dimension to describe evolution of universe are also considered as alternative theories of gravity. Interestingly, the Gauss Bonnet invariant $\cal{G}$ appears naturally in the quantum field theory regularization and re-normalization, which further leads to a second order theory even in 4-dimension in a combination of $\cal{G}$ coupled with dynamical dilaton field. This term has a non-trivial contribution in dimension $n>4$ for a constant coupling. The Gauss Bonnet dilatonic coupled term arises naturally as the leading order in the low energy effective Heterotic string theory. The Einstein Gauss Bonnet dilatonic coupled (referred as DEGB) theory works quite well in explaining inflation \cite{Neu:jcap}, as well as late-time acceleration \cite{sami:jcap} and used also to study possible resolution of initial singularity \cite{Kanti:prd, Anto:np}.
\par
Recently, some works \cite{kanti:dadhi,suma:sengup, Fomin:che,planck:cmb} have enlightened viable inflationary model based on Einstein Gauss Bonnet scalar coupling. In \cite{kanti:dadhi,suma:sengup} the scalar coupled Gauss Bonnet term alone is capable of obtaining inflationary universe without introducing slow-roll approximation. Further, slow-roll inflation potential may be in good agreement with Planck data \cite{planck:cmb} when the scalar field has a non-minimal coupling to the Gauss Bonnet term.
\par
Cosmic singularity problem has not yet been resolved irrespective of success of the aforesaid theories. However, one may avoid initial cosmic singularity introducing Euclidean wormhole in the very early universe.
Further, a mechanism of vanishing of cosmological constant was proposed by Coleman\cite{co:nu}, Baum \cite{baum:pl} and Hawking \cite{haw:pl} invoking the idea of wormhole at the Planck epoch. Wormholes are considered as gravitational instantons, which are the saddle points of the Euclidean path integral. Aforesaid wormhole represents the tunnelling of the universe through the classical singularity and this type of wormhole is classified as the Euclidean wormhole. One can also find Euclidean wormhole configuration from the solutions of the Euclidean field equations.
Lorentzian wormhole is another kind of wormhole, distinct from Euclidean wormhole. The Einstein-Rosen bridge or the Morris-Thorne wormhole \cite{morris1,morris2} are Lorentzian wormhole representing non-trivial geometry in interstellar space as a passage connecting two distinct regions in space.
\par

\par
Giddings and Strominger \cite{gs:np} first proposed a gravitational instanton as a solution of Euclidean field equations introducing axionic field in the Einstein’s theory. In our work, we specify wormhole as a configuration governed by the Euclidean field equations, wherein two asymptotic regions of Robertson walker spacetimes are connected by a tube of minimum radius of scale factor, wherein $a'(\tau_0)=0$, $a''(\tau_0) >0$ at some time $\tau_0$. Several works \cite{lee:prl,ho:plb,hall:cqg,cou:cqg,lobo:bd,nandi:bd} have been done after \cite{gs:np} to explore wormhole solutions with different fields; on the other hand, consequences of wormhole dynamics were studied extensively \cite{haw:prd,gid:npl,co:npb,fisu:plb,pol:plb,unr:prd,hawk:npb, hawk:page,kle:npb}.
The wormhole solution demands violation of null energy condition (NEC) \cite{visser:nec}, which can be minimised by introducing some geometric contribution instead of exotic fields in some theories of gravitation \cite{capp:prd,harko:prd,lobo:fR,sloan:string,met:string,Dzhu:prd,ks:fR,R:Comm}.
\par
The pre-inflationary era in cosmic evolution is yet to be explored intricately in the literature. Qualitatively, the wormhole appearing in the Planck era should reproduce an era of inflation at a later era to explain feasible cosmic evolution. The Lorentzian wormholes \cite{kanti:GB,Mehdi:GB,bhawal:GB,Mehd:Eur,ayan:euro} have been studied extensively in the DEGB theory. Recently Mehdizadeh \cite{Mehd:Eur} considered evolving Lorentzian wormhole in higher dimension in generalized Robertson Walker spacetime in Einstein Gauss Bonnet theory assuming a constraint on the Ricci scalar, which shows enlargement of a microscopic  wormhole to macroscopic size in an expanding inflationary background satisfying Weak energy condition. Again exact solution of spherical symmetric Lorentzian wormhole \cite{ayan:euro} in 4-dimension is obtained in Einstein Gauss Bonnet theory for isotropic matter, further wormhole solution is also supported by anisotropic matter assuming shape function and a particular equation of state parameter.
Substantial development of Lorentzian wormhole physics have been done in Einstein Gauss Bonnet theory, however emergence of an inflationary era from a wormhole configuration has not been well testified. So the main theme of our investigation is to study  Euclidean wormholes inducing an inflation in the DEGB theory.
\par
Euclidean wormhole solutions in higher dimension with compact internal space are presented \cite{paul:pram} including Gauss Bonnet term in the Einstein's gravity and analytic continuation of some of the Euclidean wormholes lead to the standard inflationary universe in the Lorentz spacetime. In a recent paper \cite{chew:20}, the origin of the universe has been claimed from an Euclidean wormhole in Gauss Bonnet gravity, which is based on the numerical solution of the field equations, wherein the potential of the dilaton field have been evaluated with a choice of the dilaton field introducing reverse engineering \cite{Ren:1201}.

Irrespective of above illuminating works emergence of an inflationary era from a wormhole configuration is not clearly interpreted. So our objective is to study  Euclidean wormholes inducing an inflation in the DEGB theory.
\par
The transition from a wormhole configuration in the Planck era to the inflationary era is not a smooth phenomenon because the pre-inflationary era at Planck time is determined by quantum gravity, while the inflationary era can be described by the field theory.
Cosmic evolution GUT onward is specified with proper time $t$, which is unique \cite{modak:cqg} in the classical allowed domain however time $t$ is ambiguous outside the classical allowed domain.
The space is considered Euclidean in the very early universe and in quantum field theory Euclidean time $\tau$ plays the role of time and it can probe the events inside the classical forbidden domain. Thus analytic continuation of the function from the Euclidean domain to the Lorentzian domain or change of geometry can be obtained by Wick rotation $\tau=it$, and it can be used as an important tool to study evolution from a wormhole configuration to an inflationary scenario.
\par
In this work, we investigate possible wormhole solutions in the DEGB theory in the Robertson Walker Euclidean background with a few standard potentials, and present both the analytic and numerical solutions of the Euclidean field equations. Analytic solution with a slow-roll type approximation yields Euclidean wormhole at early era (in the domain $ h(\tau)<<h_0 $) and subsequent cosmic evolution at later era $(h(\tau)>>h_0)$ shows expansion much faster than usual exponential expansion accompanied with oscillation in the asymptotic domain with                                                                                                                                                                                                                                                                                                                                                                                                                                                                                                                                                                                                                                                                                                                                                                                                                                                                                                                                 $t$ using $\tau=it$, where $h_0$ is a constant parameter and $h(\tau)$ is the Hubble parameter in $\tau$.  Further the numerical solutions satisfying all the field equation without any approximation yield dynamical scenario of the wormhole by finding the observable namely Hubble parameter $H(t)$ and the deceleration parameter $q(t)$ with a motivation to study the nature of expansion away from the throat of the wormhole.
We consider curve fit of the numerical solution of the scale factor $a(\tau)$ and transform it to $a(t)$ using $\tau=it$ in Euclidean space to the space with time $t$. The scale factor $a(t)$ is then used to find the cosmic evolution from the plot of $H(t)$ and $q(t)$ with respect to $t$. Above numerical solution obtained by using all the field equations seems to agree with the analytic solution of wormhole, since the evolution of $H(t)$ in this  numerical solution resembles with the expansion rate of external space obtained from the analytic solution of wormhole in the
Kaluza-Klein cosmology \cite{R:Comm}.
\par
Comprehensive nature of evolution of $H(t)$ and $q(t)$ from the numerical solution shows an initial collapsing era up to $t=t_i$ before encountering some unusual feature around the throat of the wormhole followed by final expansion beginning at $t=t_f$. The evolution of $H(t)$ in the domain $(t_f-t_i)$ seems unusual
in cosmic scenario. The domain $(t_f-t_i)$ shows an unusual evolution, which is almost similar to the classical forbidden domain in the analytic solution in \cite{R:Comm}. The parameter $H(t)$ approaches to a constant at $t>>t_f$, simultaneously $q(t) \rightarrow -1 $. Thus a phase transition from an Euclidean space to Hyperbolic (Lorentzian) space is obtained by the Wick rotation $\tau=it$, which yields the evolution from a wormhole configuration to an inflationary era. Though there are some unusual features near the throat, however it agrees well with the classical scenario far away from the throat of the wormhole. The potential of the dilaton field decreases to a very small value at large time. Further the null energy condition is satisfied near the throat of the wormholes in all solutions.
\par
We present the field equations in section 2. Section 3 contains a few analytic solutions introducing slow-roll type approximation in a vacuum dominated Euclidean universe. In section 4, we present numerical solutions of the field equations with some standard potentials and exponential dilaton coupling. Interpretation of the numerical solutions is given in section 5. Section 6 contains a brief discussion, finally, section 7 contains an appendix.

\section{Action with Gauss Bonnet interaction and the field equations}

We consider the action as
\begin{equation}
S=\int {d^4}x \sqrt{g}\Big[\frac{R}{2K^2} - \frac{\gamma}{2} \phi_{;\mu}\phi^{;\mu} - V(\phi) - \frac{\Lambda(\phi)}{8}{\cal{G}}\Big] + S_{m},
\end{equation}
where  $S_{m}$ is the surface term, $K$ is the inverse of Planck mass,  $\gamma$ is a constant and the Gauss Bonnet curvature is  $\cal{G} $= $R_{\mu\nu\alpha\beta}R^{\mu\nu\alpha\beta} - 4R_{\mu\nu}R^{\mu\nu} + R^2$.
Further $\Lambda(\phi)$ is the coupling of the GB term with the dilatonic field $\phi$ \cite{5:GB} and $V(\phi)$ is the potential of the  field $\phi$. In the Robertson Walker Euclidean metric
\[
ds^2 = d{\tau}^2 + a^2(\tau)\Big[\frac{dr^2}{1-\kappa r^2}+r^2 ( d{\theta}^2  + \sin^2\theta d\phi^2)\Big],
\]
the field equations from (1) yields
\begin{equation}
\frac{3}{K^2}\Big( h^2 -\frac{\kappa}{a^2}\Big)=\frac{\gamma}{2}{\phi'}^2 - V(\phi) - 3{\Lambda}'~ h \Big(h^2 -\frac{\kappa}{a^2}\Big),
\end{equation}

\begin{equation}
-\frac{1}{K^2}\Big( 2h' -\frac{2\kappa}{a^2}\Big)=\gamma \phi'^2 + {\Lambda}'' \Big(h^2 -\frac{\kappa}{a^2}\Big) + \Lambda' h \Big(2h'-h^2 +\frac{3\kappa}{a^2}\Big)
\end{equation}
and the continuity equation is.
 A prime $(^{\prime})$ stands  derivative with Euclidean time $\tau$ and a comma (,) denotes partial derivative.
 \begin{equation}
\gamma \Big(\phi''+3h\phi'\Big) =V_{,\phi} + 3\Lambda_{,\phi}\Big(h^2-\frac{\kappa}{a^2}\Big)\Big(h'+h^2\Big),
\end{equation}
where $a(\tau)$ is the scale factor,  $h(\tau)=\frac{a'}{a}$ and $\kappa (=0, \pm1)$ is the 3-space curvature Further considering (2) and (3) we get
\begin{equation}
-\frac{1}{K^2}\Big(2 \frac{{a''}}{a}+\frac{{a'}^2}{a^2}-\frac{\kappa}{a^2}\Big) = \frac{\gamma}{2}{\phi'}^2 + V(\phi)+\Lambda''\Big(\frac{{a'}^2}{a^2}-\frac{\kappa}{a^2}\Big)+2\Lambda'\frac{{a''}{a'}}{a^2}.
\end{equation}
The set of equations (2)-(5) are helpful to study evolution of the universe with knowledge of coupling $\Lambda(\phi)$ and potential $V(\phi)$.
\par
A wormhole has two asymptotic domains connected by a tube with minimum radius at the throat, so the scale factor $a(\tau)$ at the throat should have $a'(\tau_0)=0$, $a''(\tau_0) > 0$ at some
$\tau=\tau_0$, otherwise $a(\tau)$ is finite at other $\tau$. Then the field equation (2) at $\tau=\tau_0$ yields
\begin{equation}
\frac{3}{K^2}\frac{\kappa}{a_0^2}=-\frac{\gamma}{2}\phi_0'^2 +V_0,
\end{equation}
where a subscript ``$0$'' on a variable denotes the value of the variable at $\tau_0$.
So from (6) it is clear that the potential energy $V_0$  at the throat is very large as $ a_0 $ is small therein for $\kappa =1$. Again from (2) and (5), $a''(\tau)$ at the extrema is
\begin{equation}
a''_0=\frac{K^2}{3}(-\gamma\phi'^2_0-V_0) a_0 +\frac{\kappa K^2}{2}\frac{\Lambda_0''}{a_0}.
\end{equation}
The Gauss Bonnet term is assumed to be a small correction to the gravity, however the term $\frac{\Lambda_0''}{a_0}$ in (7) may be large and it may take positive or negative value. Thus to ensure lower bound of $a(\tau)$ one must satisfy
$\phi_0'^2 > V_0$ for $\gamma=-1$ and $\kappa =1$ assuming $ \frac{\Lambda_0''}{a_0}>0$.
Now we present analytic solutions with an approximation for a few  $V(\phi)$ with $\kappa = 0$ and $\gamma=-1$.

\section{Master equation to find wormhole configuration introducing slow-roll type approximation in the Euclidean field equations}
An analytic solution in closed form is non trivial, so we consider assumption to get a glimpse  of evolution in the very early universe. We present analytic solutions assuming slow-roll type approximation in the Euclidean equations.
In the early universe vacuum energy dominates over other energies and the scalar field evolves slowly towards the minimum of the potential. It was considered as a slow-roll approximation in study of inflationary cosmology.
To study the state of the universe prior to the inflationary era, solution of the Euclidean field equations (2)-(4) is necessary. The attempt by \cite{chew:20} is interesting, but not free from the ambiguous choice of dilaton field and corresponding evolution of $V(\phi)$  \cite{Ren:1201}. So we consider slow-roll type approximation in the Euclidean field equations to get a glimpse of evolution in the very early universe.
Further, the GB term is assumed to be a small correction to the gravity, then in the Euclidean section one must satisfy $\frac{1}{2}\phi'^2 << V(\phi)$, $\phi'' <<3h \phi'$, $\Lambda'h<<1$ and $\Lambda''<<\Lambda' h$. Thus we consider dominating terms in the equations (2)-(4) following \cite{gr:qc} as
\begin{equation}
\frac{3h^2}{K^2} \simeq -V-3\Lambda'h^3,
\end{equation}
\begin{equation}
\frac{2h'}{K^2} \simeq \phi'^2 +\Lambda'h^3,
\end{equation}
\begin{equation}
-3\phi'^2 \simeq \frac{V'}{h} + 3\Lambda' h^3,
\end{equation}
for $\kappa=0$ with $\gamma=-1$. Above equations (8)-(10) are identical with  \cite{gr:qc}, except the second term in the right of (8). We consider the term $\Lambda' h^3$ since the term is dimensionally same with $V(\phi)$. Now from (9) and (10) we get
\begin{equation}
6hh'= -K^2V'.
\end{equation}
Now taking derivative of (8) with $\tau$ and using (11) we get $(\Lambda' h^3)'=0$. We assume   $\Lambda' h^3 =-d $, where $d$ is a positive constant. So (8) turns to
\begin{equation}
V= - \Big(\frac{3h^2}{K^2} -3d \Big).
\end{equation}
To find evolution of $h(\tau)$ a choice of potential is necessary, so for simplicity we choose potential as $V(\phi) =V_0 \phi^n$ in (12). Now  eliminating $\phi'$ in (9) it yields
\begin{equation}
\Big(2h'+h_0^2\Big)\Big(h^2-h_0^2\Big)^{2-\frac{2}{n}} =\frac{4(-3)^{\frac{2}{n}} K^{2-\frac{4}{n}}}{V_0^{\frac{2}{n}} n^2}h^2h'^2,
\end{equation}
where $h_0^2= K^2 d$ and the constant $h_0^2$ appears from the evolution of $\Lambda'(\phi)$ and $h^3$. Further in view of (8)-(10) the term $`` 3\Lambda' h^3 "$ appears as a part of effective cosmological constant from the Gauss-Bonnet coupling. Presence of $``3\Lambda' h^3"$ in (8)-(10) or in (12) modifies effective value of dynamical cosmological constant. The equation (13) shows that $h'=0$ when $h=h_0$ for $1>\frac{1}{n}$, which yields the extremum of $h(\tau)$, so the parameter $h_0$ plays a crucial role in determining $h(\tau)$. Now (13) can be used to find $h(\tau)$ and hence the scale factor $a(\tau)$ depending on $n$. Further, exact solution of (13) is not possible in general with  arbitrary $n$. So we consider approximate solutions piece wise such that we can get an overall evolution of the universe depending on the relative value of $h^2(\tau)$ and $h_0^2$. In the forth coming section we shall present solutions with power law potentials and with an exponential potential. Before considering approximate solutions we initiate an exact solution of (13) for $\phi^2$ potential because of its simplicity.

\subsection{Wormhole configuration for  $V(\phi)=V_0\phi^2$ potential:}
We consider solution of (13) for $\phi^2$ potential. The equation (13) for $n=2$ can be simplified to the form
\begin{equation}
3h^2 h' +V_0(h^2-h_0^2)= \pm \sqrt{V_0(h^2-h_0^2)\Big[V_0(h^2-h_0^2)-3h_0^2h^2 \Big]},
\end{equation}
whose integration correspondingly yields

\begin{equation}
E[\chi, u]\pm \frac{h}{h_0}= \mp h_0 \tau,
\end{equation}
apart from a additive constant with $\tau$; where $ \chi=i \sinh^{-1}\Big(i\frac{h}{h_0}\Big) $, $u= 1-\frac{3h_0^2}{V_0}$ and $E[\chi, u]$ is elliptic function of second kind and it shows that $h=0$ at $\tau=0$, which leads to an extremum of the scale factor $a(\tau)$ in the very early universe at $\tau=0$. Above condition of the extremum of $a(\tau)$ is one of the indication for a possible wormhole solution near very small $h(\tau)$ around $\tau \sim 0$. The parameter $h_0^2$ is positive from its definition. Further the domain specified by  $h^2(\tau)<<h_0^2$ may lead to an early era of the universe, as $h=0$ at $\tau=0$. It is not possible to express $\frac{h}{h_0} $ as a function of $\tau$ in closed form from (15), so we consider the graphical plot of $\frac{h}{h_0} $ as a function of $\tau$ from (15). The right hand side of (15) is real as $``h_0 \tau"$ is real, while the left hand side  of (15) is complex function, so the real part of the left hand side of (15) leads to the variation of $\frac{h}{h_0} $ with $\tau$, which is shown in fig.1a. We can obtain the imaginary parts of (15) using  $\tau= it$ and replacing $h(\tau)$ by $h(\tau)=-iH(t)$ and we can evaluate $H(t)$ as a function of  $t$. The plot of $H(t)$ as a function of $t$ is given in fig.1b with two different values of the parameter u. Again all the curves in the plot of $\frac{H}{h_0•}$ versus $h_0 t$  in fig.1b can be fitted with $\sinh(h_0 t)$ function, and the value of $h_0$ in the continuous gray curve is $h_0$.

\par
The function $h(\tau)/h_0$  is almost linear with $\tau$ in the domain $\tau \gtrsim 6$ in the fig.1a and one may choose  $ \frac{h}{h_0}= -2m \tau + C_m$ as the solution in that domain, where $m$ and $C_m$ are constant for this curve. This lead to the solution $a(\tau)= a_0 e^{-m h_0 \tau^2 + C_m h_0 \tau}$ and under analytic continuation  from the Euclidean space to the Lorentzian space by $\tau=it$ it yields $a(t)= a_0 \exp(mh_0 t^2 + i C_m h_0 t)$. This shows an exponential expansion   much faster than usual expansion in the inflationary model accompany with oscillation of the scale factor depending on the parameter $C_m h_0$. It is not possible to extract the evolution of the universe near the domain $\tau \sim 0$ from the exact solution (15), except that the universe is at $h=0$ at $\tau=0$; which may lead to an extremum of $a(\tau)$(or may be a wormhole). Thus to know evolution in an alternative way we consider approximation in (13) to get piece wise solutions  depending on the relative values of $h^2$ and $h_0^2$. It is to note from the solution (15) that the era satisfying $h^2(\tau)<< h_0^2$  corresponds to the very early era, as the graphical solution (fig.1a) shows that $h=0$ at $\tau=0$ and at later epoch $h$ is large. We consider solution of (13) at two different regime specified by $h(\tau)$ in the forthcoming section.

\begin{figure}
\centering
\begin{subfigure}{.45\textwidth}
  \includegraphics[width=\linewidth]{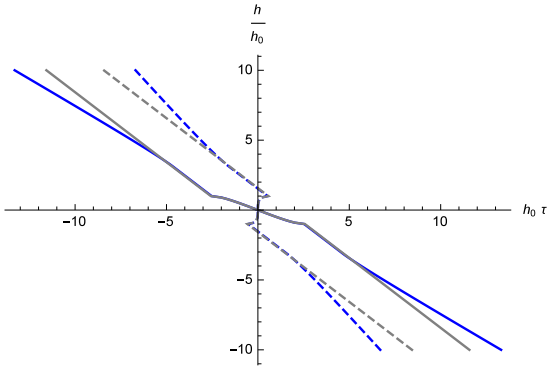}
  \caption{Fig.1a: Plot of $\frac{h(\tau)}{h_0}$ with $ h_0\tau$ from (15)}
  \label{fig:sub1}
\end{subfigure}~~~~~~~~~~~~~
\begin{subfigure}{.45\textwidth}
  \includegraphics[width=\linewidth]{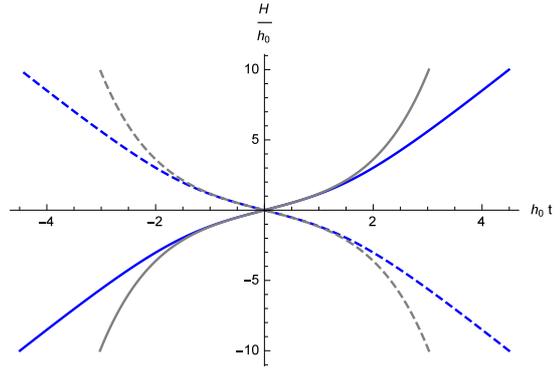}
 \caption{Fig.1b: Plot of $\frac{H(t)}{h_0}$ with $ h_0 t $ from (15) }
 \label{fig:sub2}
\end{subfigure}
\caption{The evolution of $\frac{h}{h_0}$ with $\tau$ in fig.1a is obtained for real parts of (15), while the plot of $\frac{H}{h_0}$ versus $h_0 t$ in  fig.1b is obtained for imaginary parts of (15) using analytic continuation $\tau=it$, and replacing $h(\tau) \rightarrow -i H(t)$. The parameter $u$ is taken as $u=.1$ and $u=.001$ respectively along the blue and gray curves in fig.1. Further dashed and continuous curves are used respectively for upper and lower sign of (15).}
\label{figure 1}
\end{figure}

\subsection{Evolution of the universe in the domain $h^2(\tau) << h_0^2$ (in the early era) to obtain wormhole solution:}
We consider the solution of (13) with arbitrary $n$ in the domain where  $h(\tau)$ is very small, so we choose the domain where $h^2(\tau) << h_0^2$ in the very early universe. The equation (13) then reduces to
\be
h^2h'^2 \simeq b_n(2h'+ h_0^2),
\ee
where $b_n= \frac{n^2 h_0^2}{4K^2•} \Big(  \frac{V_0 K^2}{3h_0^2•}\Big)^{2/n}$. An integration of (16) gives
\be
-C_n h \pm \frac{h}{2•}\sqrt{h^2 +C_n^2} \pm \frac{C_n^2}{2•}\ln \Big[ h+ \sqrt{h^2 +C_n^2} \Big]= C_n h_0^2~ \tau + M_n,
\ee
where $C_n^2= \frac{b_n}{h_0^2•} $ and $M_n$ is a constant. The evolution of $\frac{h(\tau)}{C_n}$ with $ \frac{2h_0^2}{C_n}\tau$ from (17) is shown in fig.2 with arbitrary n. The function $\frac{h(\tau)}{C_n}$ shows a sudden jump in fig.2 very close to $\tau=0$ and it passes $h=0$ at $ \tau=0$. Further $h(\tau)$ is very small in the early stage of evolution. Thus for small values of $h(\tau)$ in (17) we can neglect first two terms in the left side of (17) and the equation (17) yields
\be
h(\tau) = C_n\sinh\Big( \frac{2h_0^2}{C_n} \tau\Big) ~ \hbox{and}~ a(\tau)= a_0 \exp\Big[\frac{C_n^2}{2h_0^2•}\cosh\Big( \frac{2h_0^2}{C_n} \tau\Big)\Big],
\ee
wherein $h=0$ at $\tau=0$ and $a_0$ is constant. The constant $C_n$ is $C_n^2= \frac{n^2 }{4K^2•} \Big(  \frac{V_0 K^2}{3h_0^2•}\Big)^{2/n}$.
Thus (18) is a wormhole solution for arbitrary power law potential as the scale factor $a(\tau)$ is finite with $\tau$ and the radius of the wormhole at the throat is $a_0\exp(\frac{C_n^2}{2h_0^2}) $. Further analytic continuation with $\tau=it$ gives a scale factor which shows a periodic oscillation of $a(t)$ about the non-vanishing lower bound and upper maxima with respect to $t$. It is clear from above solutions that the wormhole configuration appears in the very early universe at an era $h^2(\tau)<< h_0^2$, which is independent of power law potential. Now we consider the cosmic scenario of the universe in the later epoch of $\tau$ when $h(\tau)$ is sufficiently large.

\begin{figure}
\begin{center}
\includegraphics[height=6cm,width=8cm,angle=0]{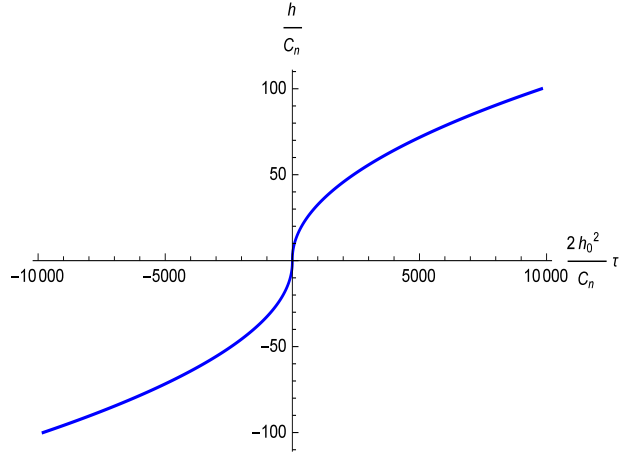}
\caption{The variation of $\frac{h}{C_n}$ versus $ \frac{2h_0^2}{C_n}\tau$ in a domain $h^2(\tau)<< h_0^2$ from equation (17) is shown by blue curve assuming $M_n=0$.}
\label{figure 2}
\end{center}
\end{figure}

\subsection{Evolution of the universe in the domain $h^2(\tau) >> h_0^2$ (in the later epoch) to study exponential type expansion :}
The evolution of the universe  when $h(\tau)$ is sufficiently large with comparison to $h_0$ in a later epoch or introducing $h^2(\tau) >> h_0^2$ in (13) with a simplification leads to
\be
h' \simeq s_n h^{2-\frac{4}{n•}} \pm s_n h^{1-\frac{2}{n•}} \sqrt{h^{2-\frac{4}{n•}}+\frac{h_0^2}{s_n}},
\ee
where $s_n = \frac{n^2}{4K^2} \Big(- \frac{V_0K^2}{3•} \Big)^{\frac{2}{n•}}$. The evolution of $h(\tau)$ from (19) is not possible with arbitrary $n$, so we consider solution with a few $n$ in following section.

\subsubsection{Solution for $V(\phi)=V_{0}\phi^2$ in a domain when $h^2>> h_0^2$ (in the later epoch):}
The equation (19) for $n=2$ in above domain  reads as
\begin{equation}
h' = \frac{1}{3}\Big[-V_0  \pm \sqrt{V_0^2-3h_0^2 V_0}\Big] ~ ~\hbox{as}~ s_2=-\frac{V_0}{3•},
\end{equation}
which gives $h'<0$ for $V_0> 3 h_0^2$, let us choose $h'= -2v<0$. Equation (20) leads to a solution
\[
a= a_0 e^{-v \tau^2},~ \hbox{which further gives}~a(t)=a_0 e^{v t^2}~ \hbox{under analytic continuation}~ \tau=it;~\hbox{ as a cosequence we }
\]
have an expansion much faster than the inflationary expansion and the Hubble parameter increases with $t$. However, equation (20) leads to a solution $a(\tau)=a_0 e^{-\frac{V_0}{6}\tau^2} e^{\pm iV_i\tau^2}$ when $V_0< 3h_0^2$  assuming $2V_i= \frac{1}{3}\sqrt{3h_0^2V_0-V_0^2} $. Thus the scale factor $a(\tau)$ is decreasing with oscillation with $\tau^2$. The amplitude of the oscillation depends on the value of $V_i$, it will be very small for very small value of $V_i$. The scale factor under analytic continuation $\tau=it$ gives $a(t)=a_0 e^{\frac{V_0}{6}t^2} e^{\mp iV_i t^2}$, which is expanding much faster than the usual exponential expansion, in addition $a(t)$ is under influence of an oscillation dependent on $V_i$. This result is almost similar with the linear part of $h(\tau)$ ($\tau\gtrsim 6$) in fig.1a obtained from (15). When $V_i$ is very small, the oscillation will be feeble and also the Hubble parameter in this case is $H(t)= \frac{\dot{a}}{a}= (\frac{2V_0}{3•}+ iV_i)t$. The imaginary part of $H(t)$
leads to an oscillation of the scale factor. The parameter $h_0^2$ appearing from the Gauss Bonnet coupling is responsible for above oscillation $a(t)$ under $V_0< 3h_0^2$.

\subsubsection{Solution for $V(\phi)=V_0\phi^{-2}$  in the domain when $h^2(\tau)>> h_0^2$ (in the later epoch):}
The solution of (19) for $n=-2$ in the domain  $h^2(\tau)>> h_0^2$ gives
\begin{equation}
\pm \frac{\sqrt{ h^4-c^4}}{h c}-\frac{h}{c}\pm 2i \Big[  E[\xi,-1]-F[\xi, -1] \Big]= \frac{h_0^2}{c}\tau +C_{11},
\end{equation}
where $\xi=i \sinh^{-1}(\frac{ih(\tau)}{c}) $,  $r= (3/V_0 K^2)^{\frac{1}{4}}$ and $c^4 r^4=h_0^2$; further $F[\xi,-1]$ and $E[\xi,-1]$ are respectively the elliptic functions of first and second kind.
A plot of $h$ (with real parts of (21)) shows linear decrease of $h$ with increasing $\tau$ at $h^2(\tau)>> h_0^2$, which are shown in fig.3a by black and blue curves respectively for upper and lower sign in (21). These curves can be fitted by the equations $ h(\tau)= - 2m_1 \tau + r_0$ away from $\tau=0$. So the scale factor is $a(\tau)= a_0 \exp(- m_1 \tau^2+ r_0 \tau)$, correspondingly $a(\tau)$ reduces to $a(t)= a_0 \exp( m_1 t^2+ i r_0 t) $  under analytic continuation $\tau=it$. Thus in the domain $h^2(\tau)>> h_0^2$ away from $\tau=0$ the scale factors expands much faster way than inflation with an accompanying  oscillation.

\begin{figure}
%\centering
\begin{subfigure}{.33\textwidth}

\includegraphics[width=\linewidth]{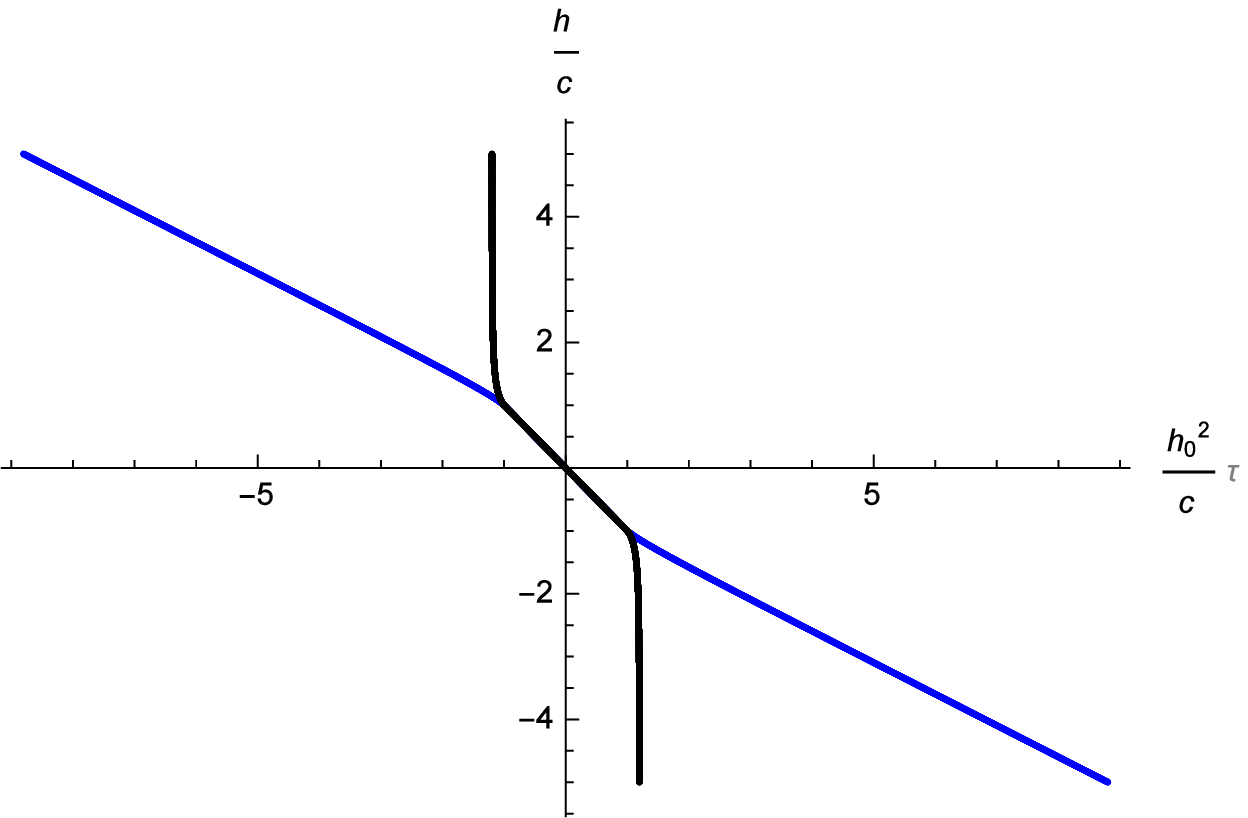}
\caption{Plot of $\frac{ h(\tau)}{c}$ versus $\frac{ h_0^2 \tau}{c}$ from (21) for $\phi^{-2}$ potential}
\label{fig:sub1}
\end{subfigure}
\begin{subfigure}{.33\textwidth}
\includegraphics[width=\linewidth]{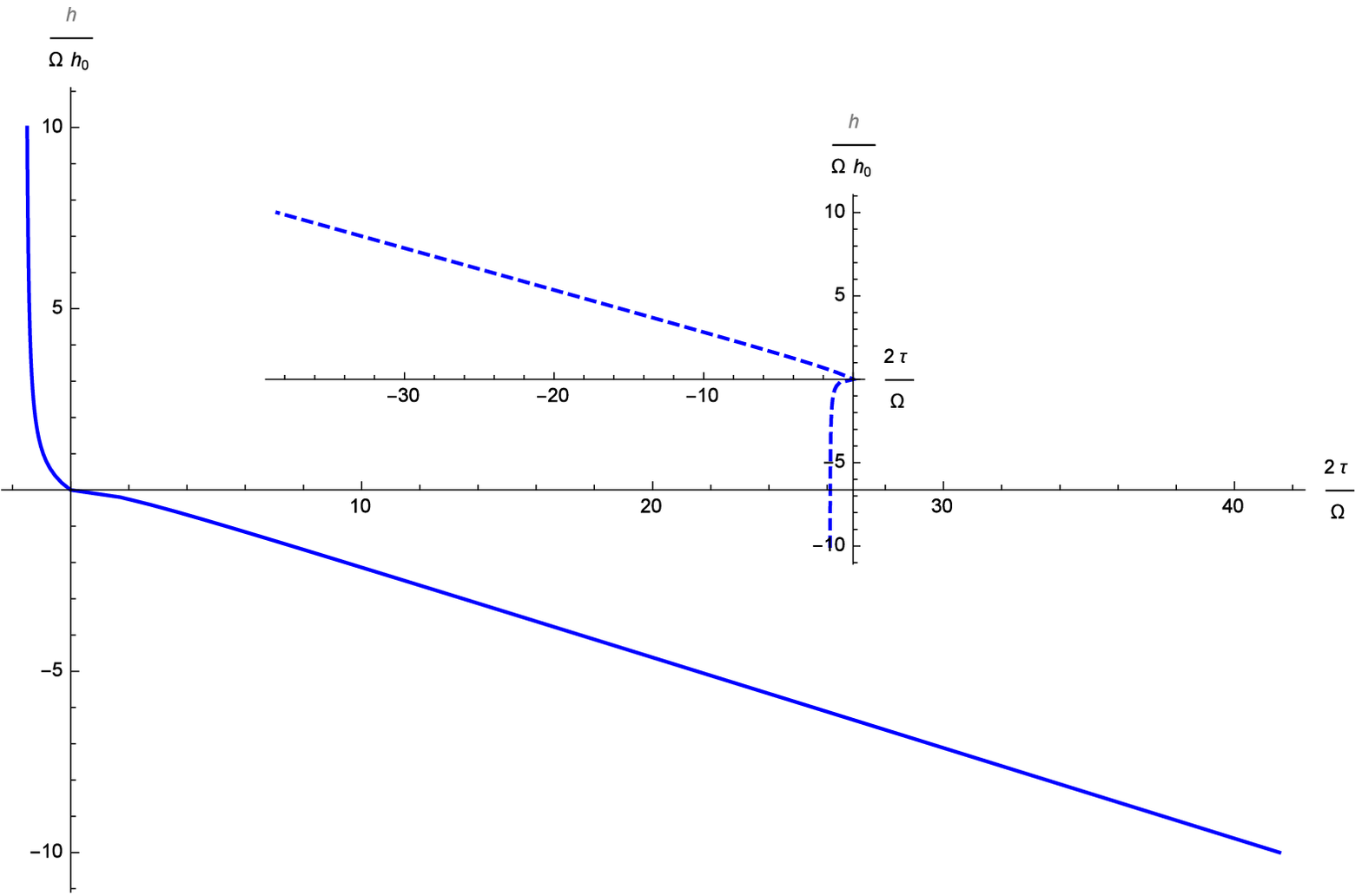} \caption{Plot of  $\frac{h(\tau)}{h_0 \Omega}$ versus $\frac{2\tau}{\Omega•}$ from (22) for $\phi^{4}$ potential}
\label{fig:sub2}
\end{subfigure}
\begin{subfigure}{.32\textwidth}
\includegraphics[width=\linewidth]{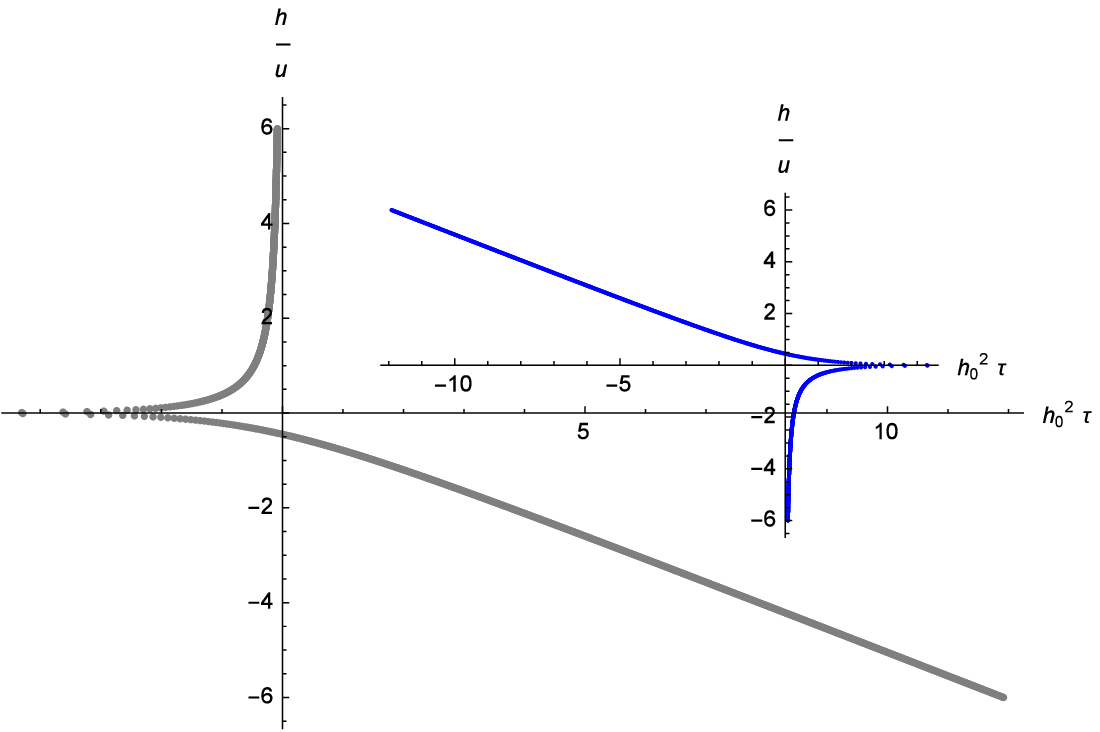}
\caption{ Plot of $\frac{ h(\tau)}{u}$ versus $ h_0^2 \tau$ from (26) for $\exp^{-\mu \phi}$ potential}
  \label{fig:sub3}
\end{subfigure}
\caption{All plots of $h(\tau)$ versus $\tau$ are considered in a regime governed by $h^2(\tau)>>h_0^2$. The evolution of  $\frac{ h}{c}$ with $\frac{ h_0^2 \tau}{c}$ for $\phi^{-2}$ potential is given in fig.3a from (21) assuming $C_{11}=0$ , wherein Blue and black curves are respectively for upper and lower sign in (21). In fig.3b the variation of $\frac{h}{h_0 \Omega}$ with $ \frac{2\tau}{\Omega}$ for $\phi^{4}$ potential is given from (22). The continuous and dashed curves  in fig.3b are respectively with upper and lower sign in (22) with $m=0$. The plot of $\frac{h}{u}$ with $ h_0^2 \tau$ is shown in fig.3c for $e^{-\mu\phi}$ potential from (26) assuming $c_2=0$.  Black and blue  curves in fig.3c are respectively with upper and lower sign in (26) with $c_2=0$.}
\label{figure 3}
\end{figure}

\subsubsection{ Solution for $V(\phi)=V_0\phi^4$  in the domain when $h^2(\tau)>> h_0^2$ (in the later epoch):}
The solution of (19) for $n=4$ in the domain $h^2(\tau) >> h_0^2$ yields
\be
-2 \bar{h}  \pm 2 \sqrt{\bar{h}^2- i\bar{h} } \pm i \ln\big[ 1 + 2i \bar{h} + 2\sqrt{-\bar{h}^2+ i\bar{h} } \Big]=\frac{2\tau}{\Omega}  + m,
\ee
where, $s_4= i\frac{4}{K} \sqrt{\frac{V_0}{3}}, $ $m$ is constant, $\bar{h}= \frac{ h(\tau)}{h_0\Omega}$ and $\Omega=\frac{K}{4} \sqrt{\frac{3}{V_0}} $. The
 function $h(\tau)$ does not allow an expression in closed form, so we present a graphical plot in fig.3b. The continuous  blue curve $h(\tau)$ shows that $\vert h(\tau) \vert$ is decreasing with increasing $\tau $ in a domain $h^2>> h_0^2$ at $\tau >0$, so we can set $ \frac{h}{h_0 \Omega}= -2p \tau -h_{01}$ in the linear part of the curve at $\tau>0$, where $p$ and $h_{01}$ are  constants. The scale factor from this $h$ gives $a(\tau)= a_0 e^{-p \Omega  \tau^2-h_{01}\tau} $ and under analytic continuation with $\tau=it$  scale factor leads to $a(t)= a_0 e^{p \Omega~t^2-i h_{01}t}$. Thus the scale factor shows an exponential expansion much faster than the usual inflationary expansion with an oscillation for $ \phi^4$ potential depending on $h_{01}$ like earlier cases.

\subsection{Evolution  of the universe for $V(\phi)=V_0e^{-\mu \phi}$ potential:}
Introducing exponential potential $ V_0e^{-\mu \phi}$ in (12) and using (11) and (9) we have
\begin{equation}
(2h'+h_0^2) (h^2-h_0^2)^2= \frac{4K^2}{\mu^2} h^2h'^2,
\end{equation}
which determines the evolution of the universe, where $\mu$ is a constant. Now to study evolution we consider solution at different regime of $h$ like earlier cases.

\subsubsection{Evolution in a domain when $h^2(\tau)<< h_0^2$ (in the early era) to study wormhole:}
The equation (23) under $h^2<< h_0^2$ reduces to
\[
h^2 h'\simeq ph_0^2 \big[ p \pm \sqrt{p^2+ h^2}\big],
\]
whose solutions is
\begin{equation}
(-p h \pm \frac{h}{2•} \sqrt{h^2+ p^2} ) \pm \frac{p^2}{2•} \ln \Big[ h+ \sqrt{h^2+ p^2}  \Big]=ph_0^2\tau +c_3,
\end{equation}
where $p^2= \frac{\mu^2h_0^2}{4K^2•}$, which is identical with (17) except the parameter $p$ ( which is $C_n$ in (17)) and it leads to wormhole solution
\begin{equation}
 a=a_0 \exp \Big[ \frac{\mu^2}{•8K^2} \cosh\Big(\frac{4K h_0\tau}{\mu•} \Big) \Big],
\end{equation}
considering dominating contribution in the second term in the left side of (24) for small $h$. This is also a wormhole solution like previous cases, but differ only with the parameters. Equivalently it gives an oscillating universe about non vanishing minima and maxima under analytic continuation $\tau=it$.

\subsubsection{ Evolution in a domain when $h^2(\tau)>> h_0^2$ (in the later epoch):}
In this condition (23) leads to
\[h'\simeq \frac{h_0^2}{u^2}\Big[ h^2 \pm h \sqrt{h^2 + u^2}\Big],
\]
accordingly its solution is
\begin{equation}
(\mp h
+ \sqrt{h^2+u^2}) + u \ln\Big[ \frac{h}{u^2 + u\sqrt{h^2+u^2}}   \Big]=\pm (h_0^2 \tau+c_2),
\end{equation}
where $c_2$ is a constant and $u^2= \frac{4K^2 h_0^2}{\mu^2}=\frac{h_0^4}{p^2}$. Obviously, (26) does not lead to a wormhole solution; however, when $h$ is sufficiently large (i.e. $h>>u$) we can estimate $h$ considering lower sign in (26) as
\begin{equation}
2h \simeq \frac{u^2}{h_0^2 \tau+c_3}-{h_0^2 \tau}- c_3, ~ \hbox{which gives}~~ a^2(\tau) =a_0^2 \Big(h_0^2\tau+c_3 \Big)^{\frac{u^2}{h_0^2}} e^{-\frac{h_0^2}{2}\tau^2} e^{-c_3\tau},
\end{equation}
 where $c_3= c_2 - u\ln u$                                                                                 . The solution (27) leads to an exponential expansion much faster than the usual inflationary expansion with an oscillation (depending on $c_3$) like earlier cases under $\tau=it$. The factor $\Big(h_0^2\tau+c_3 \Big)^{\frac{u^2}{h_0^2}} $ in $a^2$ in (27) raises problem with idea of real nature of $a^2(t)$ in the Lorentzian signature; however, we can alleviate this difficulty with choice of $ \frac{u^2} {h_0^2} $  as  $\frac{u^2} {h_0^2} =2m$ and $c_3=0$, where $m$ is an integer.

\par
In a nutshell, we have obtained wormhole solutions of identical nature independent of the exponent n for power law and exponential potential in the era specified by $h^2(\tau)<< h_0^2$ near very small $h(\tau)$. Analytic continuation by $\tau=it$ in above wormhole configuration equivalently leads to an oscillating universe between a maxima and non-vanishing minima with time $t$ and an evolution from wormhole configuration to the inflationary expansion is not possible in the domain specified by  $h^2(\tau)<< h_0^2$.
Further wormhole solutions are not allowed in the domain $h^2(\tau)>> h_0^2$, rather the universe evolved to an exponential expanding era much faster than the usual inflationary expansion accompanied by oscillations with $t$ under analytic continuation $\tau=it$. It is important to note from the above solutions  that the expansion of the universe is much faster than the exponential expansion in the domain $h^2(\tau)>> h_0^2$. So the study of slow roll parameters are not necessary. Though the pre-inflationary era shows wormhole configuration but it does not yield a viable inflationary model.
\par
The parameter $h_0^2 (=-K^2\Lambda'h^3>0)$ appears from the dynamical coupling $\Lambda(\phi)$ and $h^3(\tau)$, further the parameter $h_0^2$ and the potential $V(\phi)$ are responsible for wormhole solution. It is important to note from above analysis that the  cosmic scenario in the domain $h^2(\tau)<< h_0^2$ leads to  wormhole configuration in the very early era, which is further independent of exact form of dilaton potential. Above analytic solutions are dependent on the approximation, which may not be true in general, rather they give rise a qualitative idea of evolution in the early universe. Analytic solution with a given potential is quite complicated, so we consider numerical solutions without any approximation, except initial conditions on the variables. Now we consider numerical solutions with a few standard potentials.

\section{Numerical solution of the Euclidean field equations with $V(\phi)$ and dilaton coupling $\Lambda(\phi)$:}
Now we consider numerical solution of the field equations independent of slow-roll approximation to study wormhole for $\kappa=1$ assuming $\Lambda(\phi)$ and $V(\phi)$. Let us choose dilaton coupling $\Lambda(\phi)=\Lambda_0 e^{-\nu \phi}$ \cite{1:nep} in solving the field equations, where $\Lambda_0$ and $\nu$ are constants.  The numerical solution of (2), (4) and (5) gives us wormhole solutions with polynomial forms of potential $V(\phi)$. We present these numerical solutions graphically  using initial conditions on $\Big(a(\tau), a'(\tau), \phi(\tau), \phi'(\tau)\Big)$. The potential decays asymptotically with decaying amplitude in all cases.

\subsection{Wormhole solutions with potentials like  $V_0\phi^2$, $V_0\phi^4$ and $V_0(\phi^2-\phi_0^2)^2$:}
The numerical solution of the Euclidean field equations with potential $V(\phi)$ for scalar field yields evolution of the universe with Euclidean time $\tau$. The numerical solution gives evaluation of the scale factor $a(\tau)$ and the dilaton field $ \phi(\tau) $ graphically as a function of $\tau$. The evolution of the scale factors for polynomial potential are shown in fig.4a. The fig.4a is a superposition of plot of $a(\tau)$ for  $V_0 \phi^2$, $V_0 \phi^4$ and $V_0(\phi^2-\phi_0^2)^2$  potentials and each of these solutions have its own single minimum. The scale factor is non-vanishing with a lower bound in each solution with increasing  $\vert\tau\vert$ and finite with $\tau$, which is further almost symmetric about the minimum, so the solution represents a wormhole configuration. An important point is to note that a little variation of  initial condition does not change $a(\tau)$ versus $\tau$ significantly.
\par
The variation of potential $V(\phi)$ with $\tau$ are shown in fig.4b for aforesaid potential. In all cases of fig.4b, the potential $V(\phi)$ is maximum near the throat of the wormhole and oscillates with decaying amplitude with increasing $\vert\tau\vert$ and asymptotically decays to significantly very small, but finite value even at large $\tau$.\\

\begin{figure}
\centering
\begin{subfigure}{.45\textwidth}
  \includegraphics[width=\linewidth]{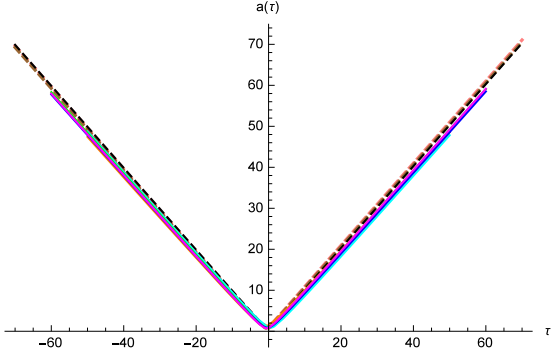}
  \caption{Fig.4a: $a(\tau)$ versus $\tau$}
  \label{fig:sub1}
\end{subfigure}~~~~~~~~~~~~~
\begin{subfigure}{.45\textwidth}
  \includegraphics[width=\linewidth]{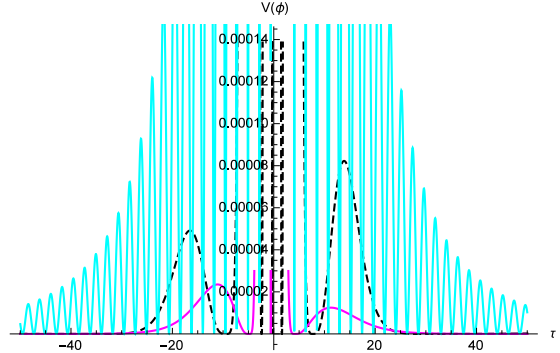}
 \caption{Fig.4b: $V(\phi)$ versus $\tau$}
 \label{fig:sub2}
\end{subfigure}

\caption{The evolution of $a(\tau)$ and $V(\phi)$ (polynomial in $\phi$) potential with $\tau$ are shown respectively in fig.4a and fig.4b  with $\gamma=-1$, $V_0=1$, $K=1$, $\kappa=1$ and $\Lambda_0=1$. Three set of initial conditions are chosen on $\Big(a(\tau),~a'(\tau),~\phi(\tau),~\phi'(\tau)\Big)$ for each potential.
The initial conditions on red, orange and cyan curves  are respectively  $(0.68, -0.18, 0.9, 0.7)$, $(1.6, .7107, .7001, 0.53)$ and $(0.88, -0.8, 1.2, 1.4)$ for $\phi^2$ potential with $\nu=0.7$ at $\tau = 0.1$.  The initial conditions on green, blue and magenta curves
are respectively $(1.15, 0.5, 0.85, -0.3)$,  $(1.1, 0.4, 0.6, -0.5)$ and $( 1.16, 0.6, 0.7, -0.2)$ for $\phi^4$ potential at $\tau=0.5$ with $\nu= 10^{-2}$, while the initial conditions on the black, brown and pink of all dashed curves are respectively $(1.15, 0.5, 0.85, 1.55448)$, $(1.1, 0.51, 0.84, -1.69036 )$ and $(1.2,  0.52, 0.86, -1.69036)$ for $(\phi^2-\phi_0^2)^2$ potential at $\tau=0.1$ with $\phi_0^2= 10^{-4}$ and $\nu=0.1$. Fig.4a is superposition of nine curves of $a(\tau)$ for above mentioned nine initial conditions, while fig.4b is superposition of  three curves of $V(\phi)$ to avoid clumsiness, viz. cyan, magenta and dashed black respectively for $\phi^2$, $\phi^4$ and $(\phi^2-\phi^2_0)^2$ potentials.}
\label{figure 4}
\end{figure}

\begin{figure}
\centering
\begin{subfigure}{.45\textwidth}
  \includegraphics[width=\linewidth]{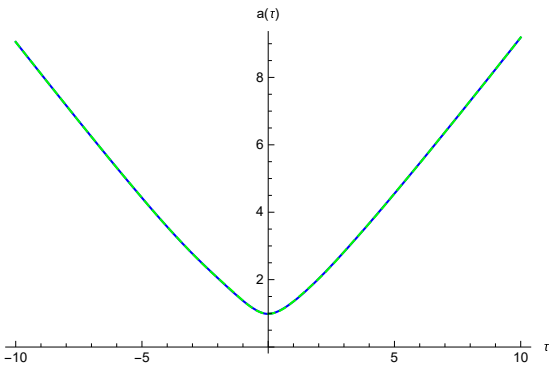}
  \caption{Fig.5a: The plot of $a(\tau)$ (cyan curve) and the corresponding fitted dashed blue curve using fit with polynomial of ``odd and even" power of  $\tau$ for $\phi^{4}$ potential.}
  \label{fig:sub1}
\end{subfigure}~~~~~~~~~~~~~
\begin{subfigure}{.45\textwidth}
  \includegraphics[width=\linewidth]{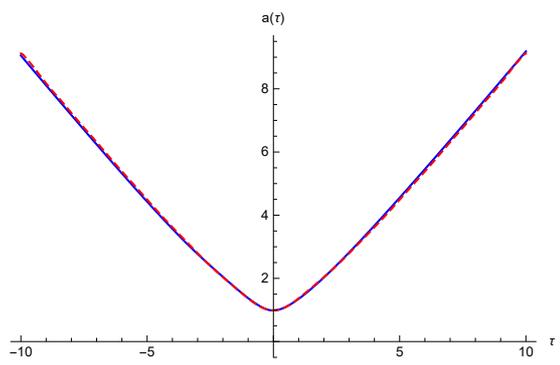}
 \caption{Fig.5b: The plot of $a(\tau)$ (blue curve) and the corresponding fitted dashed magenta curve using fit with polynomial of `` even" power of  $\tau$ for $\phi^{4}$ potential.}
 \label{fig:sub2}
\end{subfigure}
\caption{The initial condition in numerical solution of $a(\tau)$ in the cyan curve in fig.5a (or blue curve in fig.5b)  is $(a(0.5)=1.1,~a'(0.5)=-0.4,~\phi(0.5)=0.6,~\phi'(0.5)=0.5)$ with $\gamma=-1$, $V_0=1$, $K=1$, $\kappa=1$, $\nu=0.1$ and $\Lambda_0=1$.}
\label{figure 6}
\end{figure}

\begin{figure}
\centering
\begin{subfigure}{.45\textwidth}
  \includegraphics[width=\linewidth]{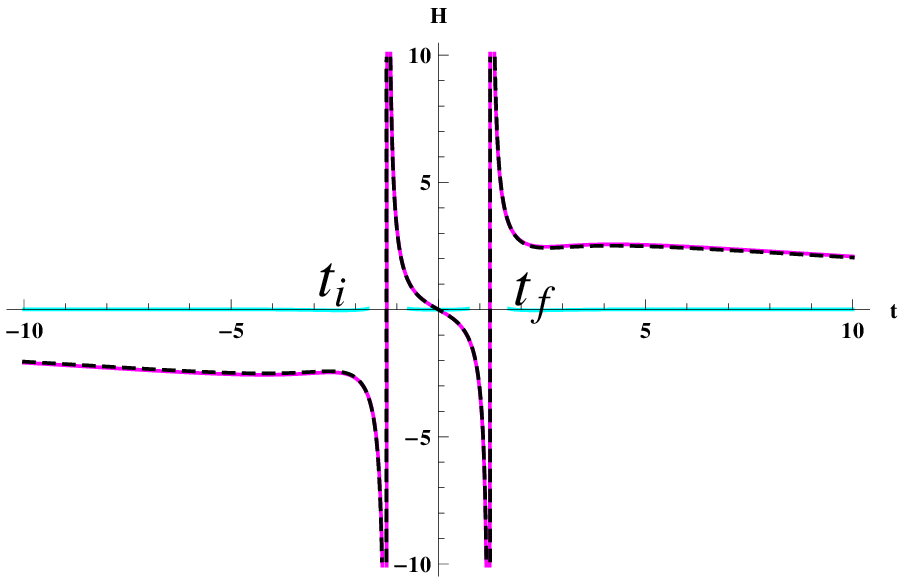}
  \caption{Fig.6a: $H(t)$ versus $t$ for $\phi^4$ potential.}
  \label{fig:sub1}
\end{subfigure}~~~~~~~~~~~~~
\begin{subfigure}{.45\textwidth}
  \includegraphics[width=\linewidth]{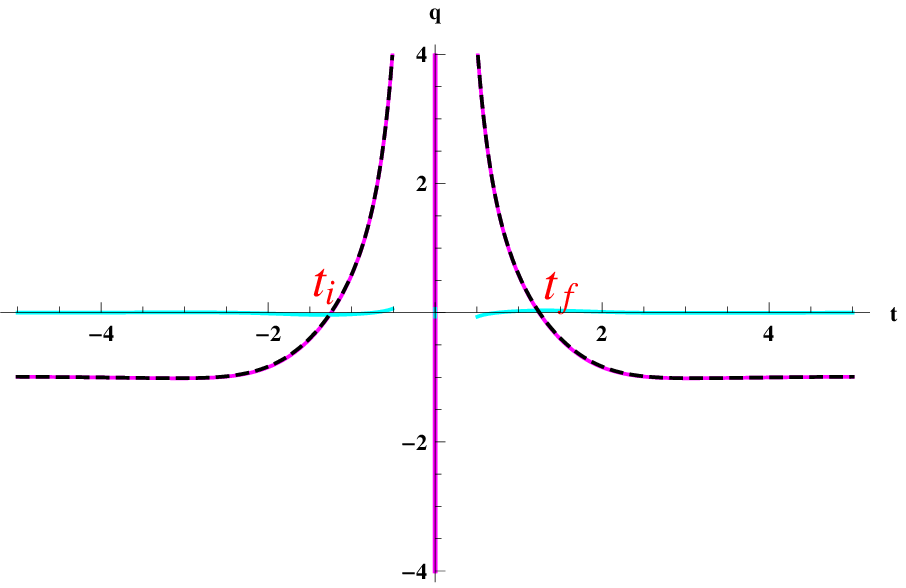}
 \caption{Fig.6b: $q(t)$ versus $t$ for $\phi^4$ potential.}
 \label{fig:sub2}
\end{subfigure}
\caption{The evolution of Hubble parameter $H(t)$ and deceleration parameter $q(t)$ with $t$ are shown respectively in fig.6a and fig.6b. Real and imaginary parts of $H(t)$ and $q(t)$ obtained from the fit of $a(\tau)$ with polynomial of ``odd and even" power of $\tau$ are shown respectively in magenta and cyan curves. Further the fit of $a(\tau)$ with polynomial of ``even" power of $\tau$ leads to real $H(t)$ and $q(t)$, which are shown by dashed curves.}
\label{figure 6}
\end{figure}

\section{Cosmic scenario of the early universe from curve fit of the numerical solution}
Now we consider consequence of the wormhole solutions, so functional form of the scale factor is necessary to study cosmic evolution of the early universe. Thus we consider curve fitting of the numerical solution of $a(\tau)$ close to the throat of wormhole for  $V_0 \phi^4$ potential. Outcome of the curve fitting is almost identical for other potentials. Hubble expansion $H(t)$ and deceleration parameter $q(t)$ are the relevant parameters at first-hand to determine cosmic evolution and we can evaluate them by using expression of $a(t)$ obtained by fit of $a(\tau)$ and analytic continuation
$\tau=it$. The plot of $H$ and $q$ as a function of $t$ yield cosmic evolution from above fitting. Two distinct expression of $a(\tau)$ may appear depending on the curve fitting in each wormhole which are discussed below.
\subsection{Cosmic evolution near the throat of wormhole by curve fit of $a(\tau)$ with a polynomial of $\tau$ for $V_0\phi^4$ :}
We consider curve fitting of the numerical solution of $a(\tau)$ within a small domain about the throat of wormhole. Two distinct expression of $a(\tau)$ may appear depending on the fit either with a polynomial of ``odd and even" power of $\tau$, or with a polynomial of ``even" power of $\tau$ in each numerical solution.  The fitted curves are shown in fig.5a and fig.5b respectively for polynomial of ``odd and even" and ``even" power of $\tau$ for  $V_0\phi^4$ potential. The expression of $a(\tau)$ using polynomial of ``odd and even" power of $\tau$ reads as
\begin{equation}
a(\tau)= \sum_{_{n=0}}^{N} b_n \tau^n,
\end{equation}
while for polynomial of ``even" power of $\tau$ is
\begin{equation}
a(\tau)= \sum_{_{n=0}}^{M} c_{2n} \tau^{2n},
\end{equation}
where $b_n$ and $c_{2n}$ are the coefficients in the polynomials, and they depend on the potential and initial conditions. The explicit expression of (28) and (29) for the potential $V_0\phi^4$ are given respectively in (30) and
(31) in Appendix-7.1. We assume the functions $1, \tau, \tau^2, \tau^3, \tau^4, .....$ with $N=30$ for $\phi^4$ potential. Further for the fit with even power of $\tau$ we choose $M =15$.
\par
The fit with only even or odd powers of $\tau$ does not reveal good fit, however a fit with $1, \tau, \tau^2, \tau^3, \tau^4, .....$ yields a better one. On the basis of the fit the metric tensor in Euclidean space $g_{ik} \propto a^2(\tau) \delta_{ik}$ are determined as a function of $\tau$ and $g_{00}=1  $; where $i, k$ runs $1, 2, 3$. Analytic continuation in the Euclidean space by the Wick rotation $\tau=it$ leads the metric tensor in terms of time $t$. However, the resulting spacetime seems to be different from a Lorentz spacetime due to the presence of real and imaginary parts of $a(t)$. The presence of odd power of $\tau$ in the fit gives rise to imaginary parts of $a(t)$ and the parameters dependent on $a(t)$ also yield real and imaginary parts. This feature is not surprising, since quantum mechanical process is dominant near the throat. In fact a wormhole configuration in general comprised of classical forbidden domain surrounded by a classically allowed region and the coordinate ``$t$" in the classical allowed domain leads to the usual time in the Lorentz spacetime. Again the coordinate ``$t$" in the classical forbidden domain changes its character from its usual notion of cosmic time, and the dynamical variables may give rise to both real and imaginary parts inside the classical forbidden domain, since ``$t$" lost its usual concept inside the classical forbidden domain, as it is governed by quantum gravity. However, a fit with polynomial of ``even" power of $\tau$ does not yield any imaginary parts either in $a(t)$ or in other variables. So we can alleviate the appearance of imaginary parts using a fit of even power of $\tau$.
\subsection{Evolution of observable parameters $H(t)$ and $q(t)$ from the fit and interpretation:}
The observable quantities  $H(t)= \frac{\dot a}{a}$ and  $q(t)= -\frac{a \ddot{a}}{{\dot{a}}^2}$ are evaluated both from (30) and (31), where an overhead dot is derivative with $t$. The plot of $H(t)$ and $q(t)$ are shown respectively in fig.6a and fig.6b for $\phi^4$ potential. The evolution of $H(t)$ and $q(t)$ in the wormhole solution is distinct from their usual real value in the cosmic evolution. However in a recent work \cite{R:Comm}, we have obtained similar evolution of $H(t)$ and $q(t)$ both from analytic and numerical solutions, but with a different action.
\par
The parameters $H(t)$ and $q(t)$ have both real (shown by magenta curves) and imaginary (drawn by cyan curves) parts close to the throat and their real part dominates over the imaginary parts away from the throat when the numerical solution is fitted with polynomial of ``odd and even" power of $\tau$. Imaginary parts of them vanish at large $t$, and asymptotically
$H(t)$ attains a constant positive value and the deceleration parameter $q(t)$ simultaneously approaches $-1$ away from the throat.\\
However, the evolution of $H(t)$ and $q(t)$ from (31) using fit with polynomial of ``even" power of $\tau$ yields a real value, which are shown in black dashed curves in fig.6. The parameter $H(t)$ approaches to a constant positive value away from the throat, while $q(t)$ simultaneously approaches  $-1$. Thus an inflationary era can be achieved from the wormhole solution by analytic continuation $\tau=it$ at time $t$ away from the throat irrespective of the fit.
\par
 It is interesting to note that the metric tensor $g_{ik}$ or $a^2(t)$ from (28), or (29) vanishes at $t= t_{i}$ and $t=t_{f}$, where $t_i=-1.3$ and $t_f= +1.3$ for $\phi^4$ potential. Further evolution of $H(t)$ in the domain $t<t_i$ is a collapsing mode, while the domain $t>t_f$ is an expanding era in the fig.6a. In between the above collapsing and expanding modes, an unusual evolution is evident in the domain $t_i< t< t_f$ surrounding the throat with respect to $t$, wherein $a(\tau)$ is well behaved with $\tau$. In fact the domain $(t_f-t_i)$ shows an unusual evolution, which is almost similar to the classical forbidden domain in the analytic solution in \cite{R:Comm}.

\par
Imaginary parts of $H(t)$ and $q(t)$, which are obtained by fit of $a(\tau)$ with polynomial of ``odd and even" power of $\tau$ are questionable from the observational ground,
however, the imaginary part of the $H(t)$ may lead to a very small oscillation  of the scale factor near the throat.
\par
Further, the potential is maximum near the throat and the fall of potential is evident from the plot of $V(\phi)$ with $\tau$. Though the potential is  decaying but it is vanishingly small even at large $\tau$. An estimate of decrease from the plot for $\phi^4$ potential with initial condition $(a(0.5)=1.1, a'(0.5)=0.4, \phi(0.5)=0.6, \phi'(0.5)=-0.5)$
( used in cyan curve in  fig.5b or blue curve in fig.5b) gives $\frac{V(\tau=0)~~}{V(\tau=10^{12})}\approx 10^{41}$,
also at later epoch it is $\frac{V(\tau=0)~~}{V(\tau=10^{22})}\approx 10^{52}$. So inflationary expansion is a consequence of the wormhole solution in the Einstein Gauss Bonnet scalar tensor theory.

\section{Discussion}

We present some wormhole solutions in the Robertson Walker Euclidean background in the dilaton Einstein Gauss Bonnet theory with a few standard potentials of the dilaton field assuming dynamical coupling $ \Lambda_0 e^{-\nu\phi} $. Analytic solutions of the wormholes are obtained introducing approximation similar to the slow-roll approximation in the Euclidean field equations in the very early universe. Wormhole is allowed in the very early era at $h^2(\tau)<<h_0^2$ (where $h_0^2=-K^2\Lambda'h^3$) with some standard potentials of the dilaton field. These Euclidean wormholes actually correspond to
oscillating universes periodic in $t$ about non-vanishing minima and maxima. Further, the solution of the field equations obtained from above mentioned slow-roll approximation in the domain $h^2(\tau)>>h_0^2$ yields expansion much faster than exponential expansion accompanied by oscillation under $\tau=it$, so it does not lead to a viable inflationary era. The above approximation also reveals that the variable ``$\Lambda'(\phi)h^3(\tau)$" appearing from the dynamical coupling acts as a part of an effective cosmological constant in the theory apart from $V(\phi)$.
\par
We also present the numerical solution of wormholes with a few standard potentials independent of slow roll approximation used in analytic solution. The plot of $a(\tau)$ versus $\tau$ shows identical wormholes independent of the form of polynomial of $\phi$ in $V(\phi)$. However, the potential evolves differently with $\tau$ for distinct $V(\phi)$, but its value is maximum near the throat and decays to a very small positive value far away from the throat irrespective of $V(\phi)$. The consequence of numerical solution of the wormholes are obtained  by curve fit  of $a(\tau)$ for $\phi^4$  potential and then transforming $a(\tau)$ to $a(t)$ introducing analytic continuation by $\tau=it$. These are then used to evaluate the Hubble parameter $H(t)$ and the deceleration parameter $q(t)$ to study the cosmic scenario of the early universe. These lead to two distinct class of variables $\{ a(t),
 H(t),  q(t)\} $ depending on the fit of $a(\tau)$ either with polynomial of ``odd and even" or ``even" power of $\tau$.
\par
The evolution of $H(t)$ (or real part of $H(t)$) shows that the domain $t<t_i$ is the initial collapsing phase, while $t> t_f$ is the final expanding phase after evolving through the throat of wormhole and the parameter $H(t)$ (or real part of $H(t)$) shows unusual evolution in the domain $t_i \leq t \leq t_f$ around the throat. This unusual regime is the classical forbidden domain.
The parameters $H(t)$ and $q(t)$ have both real and imaginary parts, when $a(\tau)$ is fitted with polynomial of ``odd and even" power of $\tau$. Further, their real parts dominate over the imaginary parts outside the classical forbidden domain and their imaginary parts vanish far away from the classical forbidden domain.
Imaginary parts of the parameters $\{ H(t), q(t)\}$ are questionable from the observational point of view.
We can alleviate the appearance of their imaginary parts  using fit of $a(\tau)$ with polynomial of ``even" power of $\tau$, though the fit of $a(\tau)$ with ``odd and even" power of $\tau$ is much better.
The plot of $H(t)$ and $q(t)$ evaluated from the fit with polynomial of ``even" power of $\tau$ are identical with the real parts of $H(t)$ and $q(t)$ obtained from ``odd and even" power of $\tau$. In all the  solutions NEC is satisfied near the throat of the wormhole.
\par
The Hubble parameter $H(t)$ approaches to a constant value far away from the throat (i.e. $t>>t_f$), and the deceleration parameter $q(t)$ simultaneously approaches to $ -1$, while the dynamical potential $V(\phi)$ approaches to very small value at $t>>t_f$. An estimate of decrease yields
$\frac{V(\tau=0)~~}{V(\tau=10^{22})}\approx 10^{52}$ for $\phi^4$ potential with initial condition given in blue curve in fig.5b (or cyan curve in fig.5a). Thus the cosmic scenario of an Euclidean wormhole leads to an exponentially expanding universe under analytic continuation by $\tau=it$. We have also obtained identical nature of $H(t)$ and $q(t)$ in Kaluza Klein theory \cite{R:Comm} inducing inflation from Euclidean wormhole configuration. So, we see that irrespective of the nature of potential or modification in the standard Einstein-Hilbert action term wormhole solutions are leading to viable inflationary era.

\section{Appendix}
\subsection{Expression of $a(\tau)$ using fit of numerical solution:}
 Expression of $a(\tau)$ using fit of numerical solution with polynomial in $\tau$:
The fit of numerical solution of $a(\tau)$ using  polynomial of odd and even power of $\tau$ within $-10\leq \tau \leq10$ for $V_0\phi^4$ potential with initial condition of fig.5 gives
\begin{multline}
a(\tau)=0.9880412 - 0.0040823 \tau +  0.4526417 \tau^2 - 0.012010 \tau^3 -
  0.0895083 \tau^4 + 0.0053538 \tau^5 \\+  0.0160261 \tau^6 - .88932998 * 10^{-3} \tau^7 -
  1.8558007 *10^{-3} \tau^8 +  0.82989 * 10^{-4} \tau^9 + \\ 1.4200181 *10^{-4} \tau^{10} -
  4.953028*10^{-6} \tau^{11} -  7.458116*10^{-6} \tau^{12} +  2.012126*10^{-7} \tau^{13} +\\
  2.769722*10^{-7} \tau^{14} -  5.755959*10^{-9} \tau^{15} -  7.4161642*10^{-9} \tau^{16} +
  1.1796636*10^{-10} \tau^{17}\\ +  1.4457887*10^{-10} \tau^{18} -  1.739694*10^{-12} \tau^{19} -
  2.0521886*10^{-12} \tau^{20} +  1.8304965*10^{-14} \tau^{21}\\ +  2.0971144*10^{-14} \tau^{22}
  -
  1.340382*10^{-16} \tau^{23} -  1.5020549*10^{-16} \tau^{24} +  6.4883410*10^{-19} \tau^{25} + \\
  7.1532807*10^{-19} \tau^{26} - 1.8660686*10^{-21} \tau^{27} -  2.0337028*10^{-21}\tau^{28} +\\
  2.4140951*10^{-24} \tau^{29} +  2.6111975*10^{-24}\tau^{30}.
\end{multline}

Data fit of numerical solution of $a(\tau)$ using polynomial of even power of $\tau$ within $-10\leq \tau \leq10$ for  $V_0\phi^4$ potential with initial condition of fig.5 gives
\begin{multline}
a(\tau)=0.9884418 + 0.4508499 \tau^2 -  0.087877 \tau^4 + 0.01536835 \tau^6 -\\
  1.7233072* 10^{-3} \tau^8 +  0.0001.2677247* 10^{-4} \tau^{10} -  6.3609441*10^{-6} \tau^{12} +\\
  2.2434738*10^{-7} \tau^{14} - 5.6716181*10^{-9} \tau^{16} +  1.0376202*10^{-10} \tau^{18} -\\
  1.3732314*10^{-12} \tau^{20} +  1.299162*10^{-14} \tau^{22} -  8.5466796*10^{-17} \tau^{24} +\\
  3.7046371*10^{-19} \tau^{26} -  9.4850915*10^{-22} \tau^{28} +  1.0828371*10^{-24} \tau^{30}.
\end{multline}

\subsection{Null energy condition:}
We can evaluate the null energy condition (NEC) by finding the sum of effective values of the energy density $\rho_{eff}$ and pressure $p_{eff}$ using the field equations (2)-(3) and the solution of the wormhole (18), which gives
\begin{equation}
\rho_{eff} +p_{eff}= \frac{4h_0^2 K^2}{C_n} \cos\Big(\frac{2h_0^2 t}{C_n}\Big).
\end{equation}
This shows that the NEC is not violated in the neighbourhood of the throat. As  $(\rho_{eff} +p_{eff})$ is oscillating, so there is periodic violation of NEC in this analytic solution. Further, the NEC for numerical solution of $a(\tau)$ can be obtained by using fit of $a(\tau)$ and analytic continuation $\tau=it$. The variation of $(\rho_{eff} +p_{eff}) $ with the cosmic time $t$ is shown in fig.7, which shows that the NEC is not violated in the neighbourhood of the throat. ($\rho_{eff} +p_{eff})$ almost attains a vanishing value in the range $-4>t>4$  and the NEC is thus satisfied in the given time domain. The imaginary parts of $(\rho_{eff} +p_{eff}) $ is very small with comparison to its real value. The imaginary part  appeared due to choice of fit of $a(\tau)$ with polynomial of both odd and even power of $\tau$, we can alleviate the appearance of the imaginary part using fit of $a(\tau)$ with polynomial of even power of $\tau$ and all the dynamical observables turned to be real function.\\

\begin{figure}
\begin{center}
\includegraphics[height=6cm,width=10cm,angle=0]{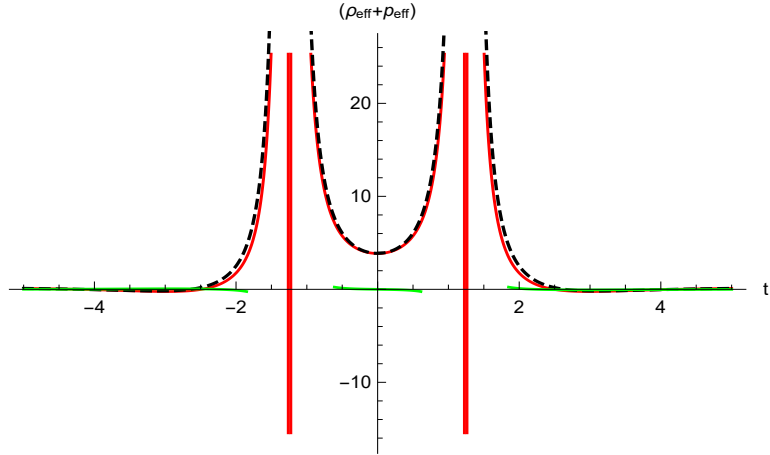}
\caption{The variation of $(\rho_{eff}+p_{eff})$ versus $t$ for $V_0 \phi^4$ potential. The red and green curves are respectively for real and imaginary parts of $(\rho_{eff}+p_{eff})$ using fit of $a(\tau)$ by polynomial of odd and even power of $\tau$ given in (30), while black dashed curve is obtained by fit of $a(\tau)$ using polynomial of even power of $\tau$ given in (31).}
\label{figure 7}
\end{center}
\end{figure}

\end{document}